**Subtractive 3D Printing of Optically Active Diamond Structures**

*Aiden A. Martin, Milos Toth[*], and Igor Aharonovich[§]*
School of Physics and Advanced Materials, University of Technology,
Sydney, P.O. Box 123, Broadway, New South Wales 2007, Australia

**Abstract**
Diamond has recently attracted considerable attention as a promising platform for quantum technologies, photonics and high resolution sensing applications. Here we demonstrate a chemical approach that enables the fabrication of functional diamond structures using gas-mediated electron induced etching. The method achieves chemical etching at room temperature through the dissociation of surface-adsorbed $H_2O$ molecules by electron irradiation in a water vapor environment. High throughput, parallel processing is possible by electron flood exposure and the use of an etch mask, while single step, mask-free three dimensional fabrication and iterative editing are achieved using a variable pressure scanning electron microscope. The electron induced chemical etching paves the way to a transformative technology for nanofabrication of diamond and other wide band-gap semiconductors.

Diamond, long considered unconquerable due to its extraordinary strength and chemical resistance, has found applications across numerous areas of science due to its unique combination of optical, electronic, chemical and thermal properties[1]. Most notably, the nitrogen vacancy luminescence center (NV) has been employed as a spin qubit, enabling the use of diamond as a platform for next generation sensing, nanopthotonic and quantum devices[2-8]. These tantalizing applications are, however, overshadowed by challenges in fabrication arising from its extraordinary hardness and chemical resistance.

At present, diamond fabrication requires cumbersome masking techniques, and ion bombardment or high power laser ablation which often causes damage and material redeposition artifacts[9-13]. Fabrication and editing of optoelectronic grade nanostructures is therefore extremely limited relative to conventional semiconductors such as silicon and gallium arsenide. Furthermore, there is no robust technique for direct-write, deterministic patterning of diamond that does not produce severe surface damage caused by ion implantation and redeposition of non-volatile, sputtered or ablated material.

Here we demonstrate a chemical approach that enables the fabrication of functional diamond structures using gas-mediated electron induced etching (Figure 1a). The method achieves chemical etching at room temperature through the dissociation of surface-adsorbed water molecules by electron irradiation in a water vapor environment. The process utilizes low energy electrons which do not cause damage through knock-on displacement of carbon, sputtering, and staining characteristic of reactive ion etching and focused ion beam sputtering techniques. High throughput, parallel processing is possible by electron flood exposure and the use of an etch mask, while single step, mask-free three dimensional fabrication and iterative editing are achieved using a variable pressure scanning electron microscope. The mask-free, electron beam induced etching (EBIE)[14, 15] is used to realize direct-write subtractive 3D printing of diamond nanostructures on inclined planes. This variant enables iterative editing and imaging of individual nanostructures. The processes are demonstrated using a standard variable pressure scanning electron microscope (SEM) making diamond nanofabrication accessible to every nanotechnology laboratory in the world. Figure 1b-c shows the steps used to direct-write etch diamond. The process involves exposure of diamond to water vapor and concurrent irradiation with an electron beam, giving rise to volatilization of carbon.[16-18]

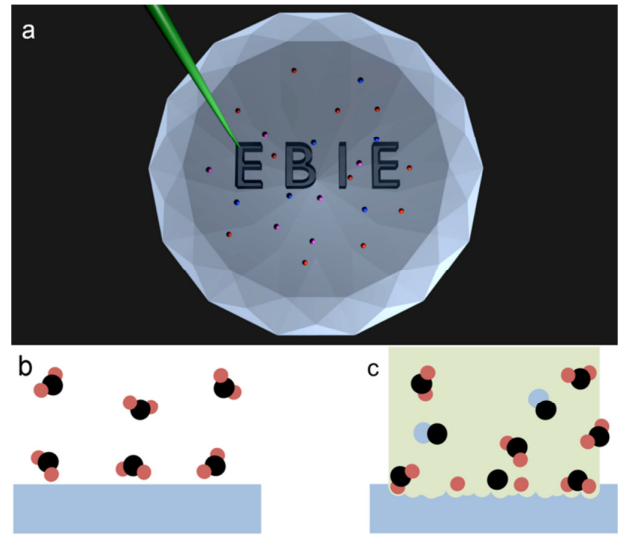

**Figure 1.** Schematic illustrations of $H_2O$ mediated electron beam induced etching. **a** Direct-write subtractive printing of diamond. **b-c** Volatilization of diamond by electron exposure in a gaseous H2O environment.

To demonstrate the applicability of EBIE to device fabrication, we start by fabricating a pillar from a single crystal diamond using an etch mask. Pillars are used as antennas that enhance light extraction from the embedded emitters, particularly of high refractive index semiconductors. The EBIE process is shown schematically in Figure 2a-c. The mask must either absorb the incident electrons or prevent H2O from adsorbing to the diamond substrate. Here we use a silica mask to prevent low-energy (2 keV) electrons from penetrating into underlying regions of diamond. The resulting pillars (Figure 2d) have high aspect ratios and straight side-walls, making them ideal for photonic applications. The minimum pillar diameter is ultimately limited by the diameter of the interaction volume of a delta function electron beam, which scales super-linearly[19] with electron energy. In diamond, it is equal to ~ 19 nm at 2 keV, and ~ 9 nm at 1 keV, as shown in Figure 2e.[20] Nanostructures can therefore be fabricated with high resolution using the correct combination of mask diameter and electron energy. On the other hand, micron sized depths can be achieved, enabling high aspect ratio geometries.

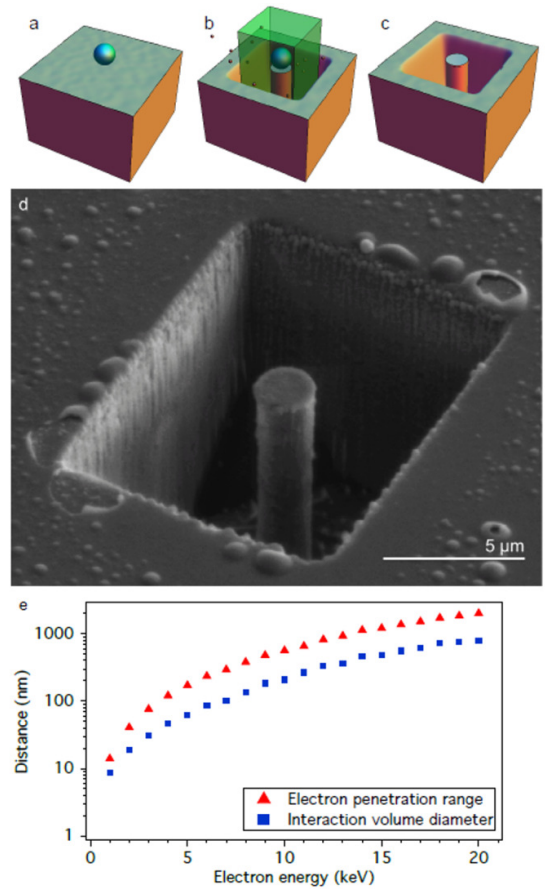

**Figure 2.** Diamond pillar fabricated by mask-based EBIE. **a - c** Schematic illustration of pillar fabrication by $H_2O$ mediated EBIE. **a** Silica bead on a diamond surface. **b** Diamond volatilization achieved by scanning a 2 keV electron beam over a rectangle repeatedly in a H2O environment. The silica bead acts as an etch mask that prevents the electrons from reaching the diamond surface. **c** Final pillar geometry after the silica bead was removed from the substrate. **d** Electron image of a pillar fabricated in single crystal diamond by H2O-mediated EBIE using the process shown in a - c. **e** Depth and diameter of the electron interaction volume that contains 90% of the energy deposited into diamond, plotted as a function of electron energy. The values were calculated using a Monte Carlo model of electron-solid interactions.

Optical properties of the pillars are shown by the fluorescence and Raman scattering data in Figure 3. PL spectra were recorded using a confocal microscope with a 532 nm excitation laser. The PL intensity of the pillar (Figure 3a) shows a two fold increase over the neighboring, at, unprocessed region of diamond (under identical PL collection conditions).
Raman spectroscopy (Figure 3b) shows no evidence of graphitic inclusions in the irradiated area with the first-order diamond peak positioned at 1332 cm$^{-1}$ and FWHM of ~ 3.6 cm$^{-1}$, consistent with the Raman signature of pristine, single crystal diamond[21, 22].

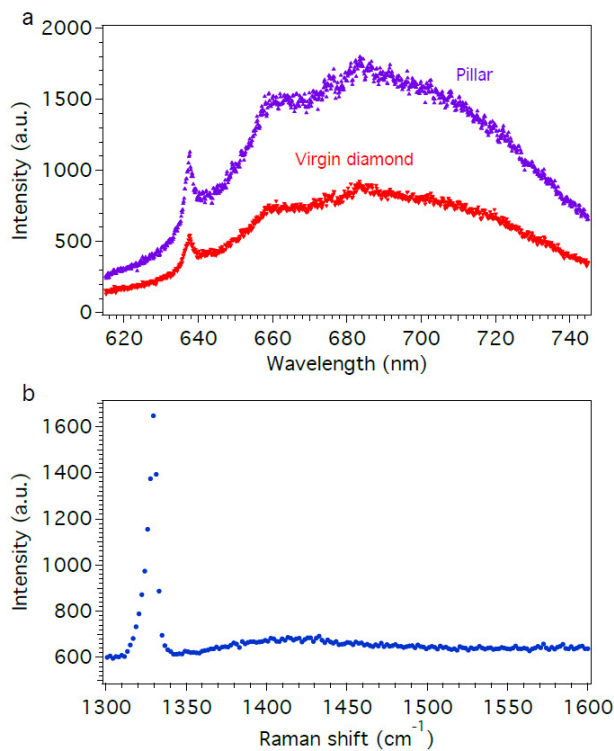

**Figure 3.** Optical quality of a diamond pillar fabricated by EBIE. **a** Photoluminescence spectra of the pillar and an adjacent, unprocessed region of diamond. **b** Raman spectrum of the pillar showing the absence of defects generated by EBIE.

Next, we demonstrate the unparalleled capability of EBIE for mask-free editing of inclined diamond surfaces. Editing of multiple inclined facets is nearly impossible by mask-based processing techniques, including electron- and photo-lithography. To demonstrate the three dimensional capability of writing on inclined surfaces, we patterned the letters `UTS' and `NANO' into individual microparticles (Figure 4) simply by tracing out the letters using an electron beam as shown schematically in Figure 1d. Etching was carried out using a 20 keV electron beam, while charging was stabilized using a low vacuum (13 Pa) environment of $H_2O$. The letters are clearly visible in SEM images (Figure 4a), while atomic force microscope (AFM) maps of the `UTS' logo show line widths and depths of ~ 100 nm (Figure 4b). The letters `NANO' were written intentionally across three diamond (111) facets, showing the ability of EBIE to edit three dimensional, inclined nanostructures. Figure 4c shows an individual diamond microparticle with visible (111) facets and Figure 4e shows the word `NANO' imprinted in the crystal, with the letters `NA', `N' and `O', occupying all three (111) planes, respectively. PL measurements recorded from the diamond microparticles exhibit strong fluorescence, confirming that the etch process does not destroy optical properties and material functionality. The mask-less patterning approach is particularly attractive for generation of high resolution microfluidic channels in microdiamond crystals, in a close proximity to optical emitters.[23, 24]

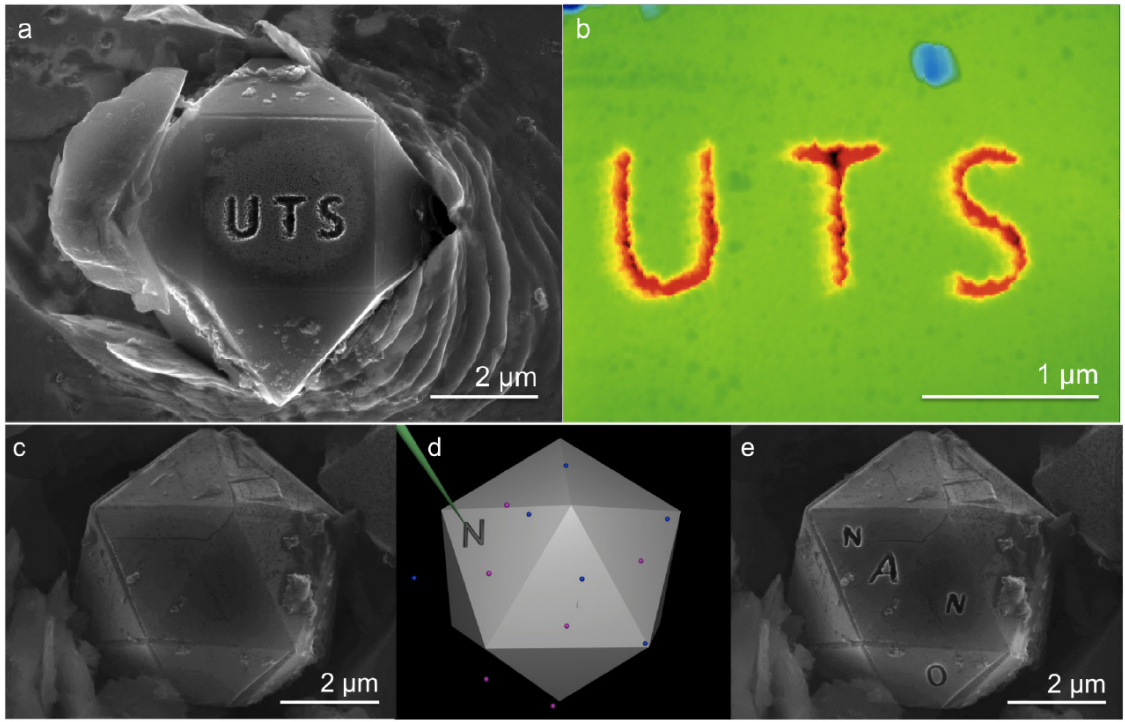

**Figure 4.** Beam-directed editing of Si-doped diamond micro-particles. **a** SEM image of the symbol `UTS' patterned by $H_2O$ mediated EBIE on the 110 plane of a single diamond micro-particle embedded in platinum. **b** AFM image of the symbol `UTS' shown in a (depth of each letter ~ 100 nm). **c** SEM image of a diamond micro-particle. **d** Schematic illustration of the process used to pattern the micro-particle shown in c. Each letter of `NANO' was patterned individually using H2O mediated EBIE on three different 111 faces of diamond. **e** SEM image of the microparticle shown in c after the letters `NANO' were patterned by EBIE.

The potential of EBIE exceeds that of traditional etching techniques for wide bandgap semiconductors. For instance, a combination of EBIE with cathodoluminescence analysis techniques may enable probing of the spectroscopic properties of nanostructures while the etch parameters are modified during fabrication. Alternatively, substrate tilting can enable fabrication of undercut structures that are currently not available in diamond. Finally, the EBIE method will be pivotal for realizing hybrid devices when direct sculpting of a nanostructure is required to achieve close proximity with an external cavity or metallic nanostructure[25].

We have demonstrated a novel, promising approach to pattern and sculpt optically active diamond structures using two variants of H2O-mediated electron induced chemical etching: a scalable, mask-based lithographic approach, and an extremely versatile, direct- write editing process. For the first time, direct 3D writing is realized on various facets of a single microparticle. PL and Raman analysis were used to show that the unique optical properties of diamond are maintained and no graphitization occurs. By leveraging the advanced functionalities provided by a conventional SEM in conjunction with EBIE a modification to existing devices and direct nanofabrication for rapid prototyping is enabled. The realized electron induced chemical etching paves the way to a transformative technology for nanofabrication of diamond structures and other wide band-gap semiconductors. It is the first step towards rendering 3D single crystal diamond geometries for high performance photonic, sensing and quantum devices.

**Acknowledgements**
This work was partly funded by FEI Company. A.A.M. is the recipient of a John Stocker Postgraduate Scholarship from the Science and Industry Endowment Fund. I.A. is the recipient of an Australian Research Council Discovery Early Career Research Award (Project Number DE130100592).

Correspondence to: milos.toth@uts.edu.au  or igor.aharonovich@uts.edu.au